# Molecular-Induced Chirality Transfer to Plasmonic Lattice Modes

E.S.A. Goerlitzer[1,§,*], M. Zapata-Herrera[2], E. Ponomareva[3], D. Feller[3], A. Garcia-Etxarri[4,5], M. Karg[3], J. Aizpurua[2,4,*], N. Vogel[1,*]

[1] Institute of Particle Technology, Friedrich-Alexander University Erlangen-Nürnberg, Cauerstraße 4, D-91058, Erlangen, Germany

[2] Materials Physics Center CSIC-UPV/EHU, Paseo Manuel de Lardizabal 5 20018, Donostia-San Sebastián, Spain

[3] Institut für Physikalische Chemie I: Kolloide und Nanooptik, Heinrich-Heine-Universität Düsseldorf, Universitätsstr. 1, Düsseldorf, D-40225 Germany

[4] Donostia International Physics Center (DIPC), Paseo Manuel de Lardizabal 4 20018, Donostia-San Sebastián, Spain

[5] IKERBASQUE, Basque Foundation for Science, Maria Diaz de Haro 3, 48013 Bilbao, Spain

[§] Current address: NanoPhotonics Centre, Cavendish Laboratory, University of Cambridge, Cambridge CB3 0HE, U.K.

* Corresponding authors: eric.goerlitzer@fau.de, aizpurua@ehu.eus, nicolas.vogel@fau.de

## Abstract

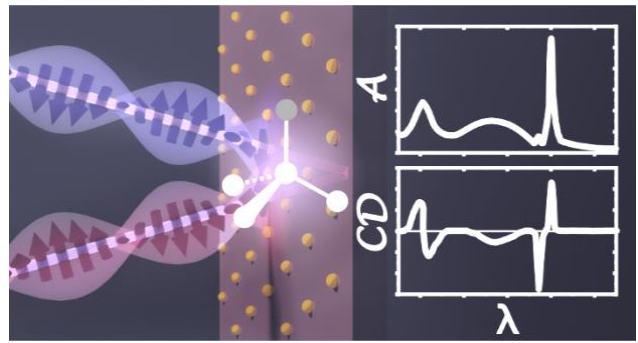

Molecular chirality plays fundamental roles in biology. The chiral response of a molecule occurs at a specific spectral position, determined by its molecular structure. This fingerprint can be transferred to other spectral regions via the interaction with localized surface plasmon resonances of gold nanoparticles. Here, we demonstrate that molecular chirality transfer occurs also for plasmonic lattice modes, providing a very effective and tunable means to control chirality. We use colloidal self-assembly to fabricate non-close packed, periodic arrays of gold nanoparticles, which are embedded in a polymer film containing chiral molecules. In the presence of the chiral molecules, the SLRs become optically active, i.e. showing handedness-dependent excitation. Numerical simulations with varying lattice parameters show circular dichroism peaks shifting along with the spectral positions of the lattice modes, corroborating the chirality transfer to these collective modes. A semi-analytical model based on the coupling of molecular and plasmonic resonances rationalizes this chirality transfer.

# 1 Introduction

Chiral plasmonics is among nanophotonics' most intriguing and interdisciplinary fields and is on the leap towards real-life applications.[1–11] Chiral molecules are abundant yet crucially important in nature.[12] The handedness of a molecule underpins its function in biological systems, despite having an otherwise *identical* chemical nature as the enantiomeric counterpart. On a larger length scale, encoding chirality into plasmonic materials has enabled the manipulation of electromagnetic fields and thus tailoring light-matter interactions,[1–4,6,8–11] and enhancing the sensitivity of chiral molecule detection,[2,5–7] with the potential to enable early disease detection in tissues.[13] The potential to use chiral nanophotonic sensors to enhance molecular signals is appealing. However, the accurate measurement of molecular circular dichroism (CD) spectra in the ultraviolet region requires sensitive and expensive instrumentation.[14] Furthermore, the fabrication of suitable plasmonic systems for such chiral detection is complex and requires sophisticated infrastructure.[15–18]

Chiral plasmonic metamaterials exhibit much larger CD signals compared to chiral molecules.[14] The dominant chiral signal of such metamaterials might screen the intrinsic chiral information from the molecule, thus necessitating background subtraction and alignment.[19–23] These difficulties can be overcome by using racemic metamaterial structures, which do not exhibit an overall intrinsic far-field CD signal and thus do not shield the molecular signal.[19,24–28]

A conceptually much simpler alternative can take advantage of plasmonic resonances in isotropically achiral nanoparticles to detect the handedness of a molecular analyte. Such nanoparticles can become optically active by a chirality transfer from the molecules in their vicinity (see **Figure 1**a, green line and Supporting Figure 1).[17,22,29–32] A particular appealing aspect of this mechanism is that the observation of chirality occurs at the plasmonic resonance in the visible range, while the initial molecular signal can be located at lower wavelengths, i.e., in the UV.[33] However, this transfer may come at the cost of losing some information of the molecular CD spectrum as, in general, helicity is not preserved by plasmonic nanoparticles. This molecular chirality transfer has so far solely been demonstrated for localized surface plasmon resonances (LSPRs).[34–42]

The chirality transfer between molecules and LSPRs of plasmonic structures[17,22,23,29–32] is complex and remains subject to current research.[43–50] Using individual nanoparticles seems straightforward, but shows certain limitations. Single nanoparticles show relatively weak near-fields in their proximity and broad resonance features. These near-fields can be enhanced in hotspots formed by dimers or chains of nanoparticles to improve the induced chirality. However, their mode volume remains strongly confined to a small region.[34–42] Surface lattice resonances (SLRs) may provide solutions to these drawbacks. SLRs occur in periodic plasmonic nanoparticle arrays and result from electromagnetic coupling of LSPRs and diffractive modes. This coupling reduces losses, leading to sharp, spectrally tunable modes.[51–56] Characteristically, the enhanced near-fields in SLRs[51–53,55] are much more delocalized throughout the volume between the particles as compared to that in LSPRs.[57,58]

Here, we demonstrate the molecular chirality transfer to lattice resonances, using arrays of spherical gold nanoparticles, as the simplest possible realization of such systems. We hypothesize that randomly arranged molecules covering these substrates (**Figure 1**a) can successfully couple with the achiral plasmonic particles in the array. Thus, both LSPRs and SLRs in such a 'hybrid' system will show an induced chirality (**Figure 1**b, blue line) if the chiral molecules surround them. This demonstration of chirality transfer to collective plasmonic modes opens possibilities to harvest the attractive properties of lattice resonances, like spectral tunability and narrow features, towards molecular chirality detection.

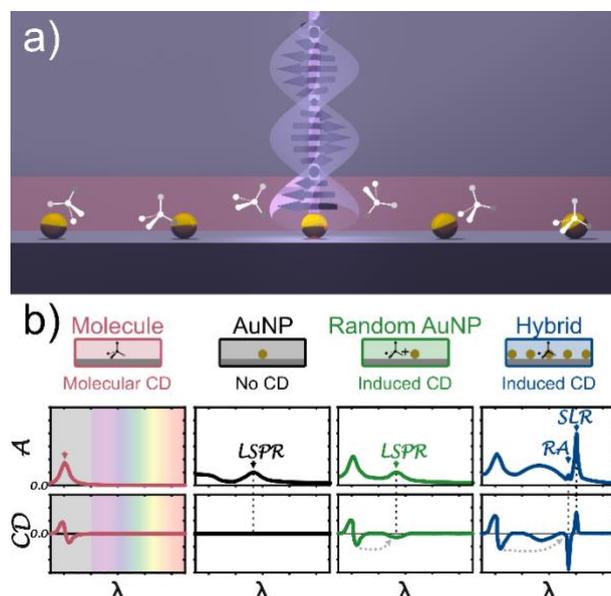

**Figure 1. Hypothesis of induced chirality in individual achiral plasmonic gold nanoparticles and their arrays.** a) A chiral film (red) consists of randomly oriented chiral molecules embedded in a thin layer of polymer (PMMA). This chiral film is sandwiched between a glass substrate and immersion oil of similar real refractive indices. b) This configuration exhibits circular dichroism (CD ≠ 0) around the molecular resonance as shown in the schematic spectra (red). This means the absorption is different for the left and right-handed circularly polarized light. Randomly arranged achiral gold nanoparticles added to a chiral film can show induced chirality at the LSPR (green). An assumed array of spherical gold nanoparticles will excite, in addition to a broad LSPR, also sharp surface lattice resonances (SLRs) and Rayleigh anomalies (RA). In this work, we extend the effect of induced chirality to lattice modes. We assume these coherent resonances can also show optical activity when chiral molecules are present in the nanoparticle arrays. Besides the LSPR, we could see that the RAs and SLRs show non-zero CD signals (blue). No actual data, schematic representation for illustrative visualization.

## 2  Results

### 2.1  Design of experimental system

We experimentally realise the proposed systems (**Figure 1**a) to examine our hypothesis (**Figure 1**b). We fabricate large-area nanoparticle arrays on a substrate via colloidal self-assembly[58,59] (**Figure 2**a). We use core-shell (CS) microgels with plasmonic NP cores (AuNPs; $D_{TEM}$ = 99 nm) surrounded by *soft* poly(N-isopropylacrylamide) (PNiPAm) shells (bulk diameter $D_{h,\,core\text{-}shell}$ = 367 nm). These CS microgels are spread at the air-water interface until the interface is fully covered and a homogeneous monolayer of CS microgels in hexagonal arrangement is formed. The monolayer is then transferred to a glass slide (**Supporting Figure 2**). After the first deposition, the remaining interfacial CS monolayer relaxes into a hexagonal array with increased lattice constant $D_{hex}$, which again can be transferred to a solid substrate.[58,59] After transfer, the organic PNiPAm shell is removed by oxygen plasma, revealing the hexagonal array of AuNPs on the glass substrate (**Figure 2**b). Using this method, we prepare samples with two different lattice constants, where the cores are separated by $D_{hex}$ ~ 450 nm and $D_{hex}$ ~ 550 nm (Exemplarily shown for $D_{hex}$ ~ 450 nm in **Figure 2**b). Due to fabrication imperfections and a reduced coupling strength between the plasmonic resonance and the Rayleigh anomaly (RA), the latter results in broader SLRs.[58]

As the chiral molecule, we use Riboflavin embedded in a poly(methyl methacrylate) (PMMA) film to allow comparison with previous experiments.[35] We modify the established protocol (details in the experimental section) to create a chiral polymer film of ~ 200 nm thickness. The side view images (**Figure 2**d) reveal that the AuNPs remain assembled on the substrate. We produce a set of two reference samples, one with arrays of AuNPs in plain PMMA films without Riboflavin, and a second one having a Riboflavin-containing PMMA film without the AuNPs. All substrate films were covered with immersion oil and a cover slide (**Supporting Figure 2** and **Supporting Figure 3**) to match the refractive index (RI) and thus effectively excite SLRs.[51–55]. The homogeneous refractive index environment also prevents scattering, so the surface roughness of the PMMA film can be ignored. The large area covered by the AuNP array allows us to measure the samples in a commercial CD spectrometer (Jasco J815a) under normal illumination with a custom build sample holder (**Supporting Figure 3**).

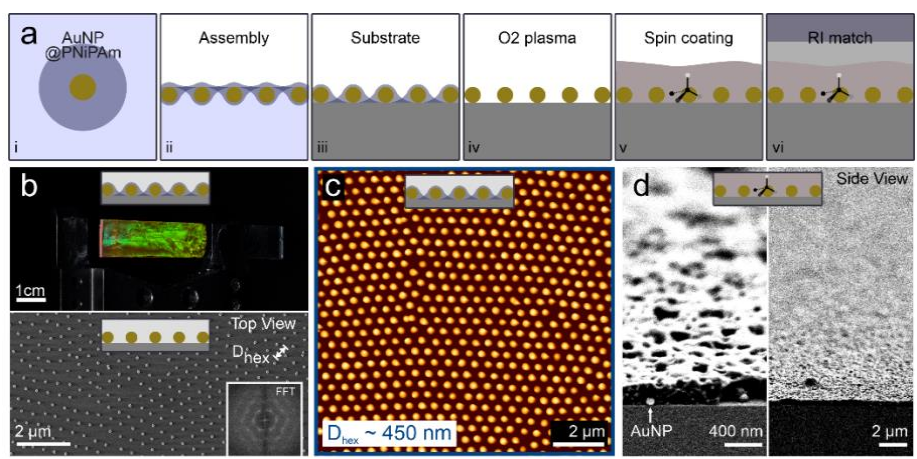

**Figure 2. Realisation of the experimental system to investigate induced chirality in surface lattice resonances.** a) Core-shell microgels (99 nm AuNP @ 367 nm PNiPAm) were assembled at the air-water interface and transferred to glass substrates. The organic shells of the resulting monolayers were removed, revealing the hexagonal arrays of AuNPs. A ~200 nm PMMA film with 40 mM Riboflavin was spin-coated on top. The RI environment was matched with immersion oil and a glass cover. b) Top: Photograph of a typical sample after the transfer of the interfacial colloidal monolayer to the solid substrate with structural coloration arising from the periodic nature of the two-dimensional array. Bottom: SEM image of the arrays of AuNPs taken after shell removal. The fast Fourier transformation (FFT) (inset) reveals the hexagonal symmetry of the assembly. c) Representative AFM image of a sample with $D_{hex}$ ~ 450 nm. Note that the organic shells are still present in this sample. d) Side-view SEM images at two different magnifications of a sample with arrays of AuNPs and a chiral film on top (without immersion oil).

## 2.2 Experimental demonstration of molecular chirality transfer to lattice modes

We optically characterise the prepared samples by illumination with circularly polarized light at normal incidence. We measure absorption under left and right circularly polarized (LCP, RCP) illumination in transmission mode to obtain the CD spectra. First, in **Figure 3**a and b we show the absorption and CD spectra for all references, the pure chiral Riboflavin molecules embedded in PMMA (red line), and two AuNP array samples without any chiral molecules (black and grey lines). The bare molecular film shows spectral features between 350 and 500 nm, both in absorption and in CD. Note that the used glass substrate is not transmissive below 350 nm. Our measured molecular signals are comparable in spectral position and intensity with the previous study of induced chirality using the same molecule.[35] The bare AuNPs arrays (without molecules) show two resonances each, a LSPR located at around 575 nm and SLRs with resonances at ~630 nm and ~720 nm for the samples with $D_{hex}$ ~ 450 nm (SLR1) and $D_{hex}$ ~ 550 nm (SLR2), respectively (see also **Supporting Figure 4**). Assuming that the SLRs result from coupling to the <1,0> diffraction mode, we expect Rayleigh anomalies at 592 nm and 724 nm, respectively, for the two interparticle spacings ($RA_{<1,0>} = \sqrt{3/4} \cdot D_{hex}$, **Supporting Figure 5**).[51–56] The spectra agree with previous results obtained from periodic arrays of similar sized, spherical AuNPs.[58,59] Note that the broadening of the SLRs arises from fabrication imperfections typical of self-assembled structures[57,60–62] and limitations in coherency and collimation of the commercial CD spectrometer.[56] As anticipated, for achiral, plain arrays of AuNPs, there are no signatures of chirality present in the CD spectra (black and grey lines in **Figure 3**b, see also **Supporting Figure 4**).

The combination of chiral molecules and achiral AuNP arrays ('hybrid' system) changes the chiroptical response. **Figure 3**c and d show the spectra for these hybrid systems (blue and yellow lines). The absorption spectra of the hybrid systems show the spectral features of the individual constituents (Figure 3c). In particular, both samples show a common molecular feature at 350-500 nm and SLRs at ~645 nm (SLR1) and ~693 nm (SLR2), for the lattice parameters $D_{hex}$ ~ 450 nm and $D_{hex}$ ~ 550 nm, respectively. In contrast, the CD spectra are seemingly not a mere addition of the two component signals (**Figure 3**d). The CD signals in the molecular region change sign and shape, presumably because interactions with the AuNPs prevent the preservation of the helicity.[48] Importantly, the CD spectra show several additional features not observed in the reference systems. Positive CD signals at ~575 nm, corresponding to the position of the LSPRs (grey arrow), are found in both samples and result from molecular chirality transfer to the LSPRs, corroborating results from literature.[35]

The sample with $D_{hex}$ ~ 450 nm additionally shows a CD peak at ~650 nm. This positive CD peak nearly coincides with the spectral position of the SLR1 peak (**Figure 3**d, blue arrow), indicating that the molecular chirality signal is transferred to the SLR of the array. Corroborating this interpretation, the CD peak shifts along with the SLR for the sample with a larger lattice spacing ($D_{hex}$ ~ 550 nm) to ~ 700 nm, which spectrally coincides with the SLR2 (**Figure 3**d, yellow arrow). Together, these observations provide experimental demonstration of plasmon-induced chirality transfer to SLRs.

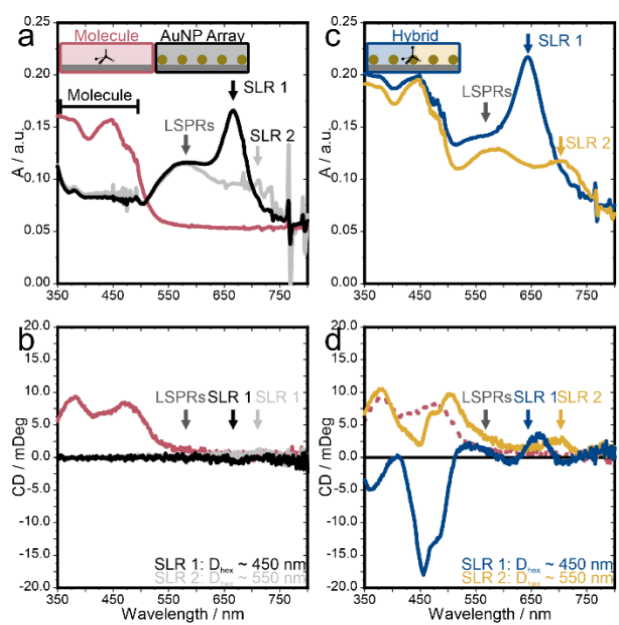

**Figure 3. Optical response of isotropic, achiral AuNP arrays embedded in plain and Riboflavin-loaded PMMA films to circularly polarized light**. a) Mean absorbance and b) circular dichroism

spectra of two reference systems with only the Riboflavin/PMMA layer (red line) and only the AuNP arrays (black and grey lines; $D_{hex}$ ~ 450 nm and ~ 550 nm). Only the Riboflavin-loaded film shows chirality, while the plasmonic arrays remain achiral at the positions of the LSPRs and SLRs. c) Mean absorption and d) circular dichroism spectra of the two hybrid samples with AuNP arrays embedded in the chiral Riboflavin-loaded film. The absorption spectra appear as additions of the individual contributions. Both samples show CD peaks at the positions of the respective surface lattice resonances (SLR1 and SLR2, indicated by arrows). Black and grey lines for $D_{hex}$ ~ 450 nm and ~ 550 nm respectively; red, dashed line shows molecular CD spectra.

## 2.3 Theoretical description of the plasmon induced chirality transfer to surface lattice modes

We systematically investigate the chirality transfer to SLRs via full-wave electromagnetic simulations (COMSOL Multiphysics®).[63] We set up an array of gold nanospheres (D = 90 nm) positioned at the bottom of a 200 nm thick chiral film, where the chiral response is introduced as specified below. We consider a square array of spherical gold nanoparticles with a variable lattice spacing $D_{Latt}$. The surrounding environment is assumed to be a homogenous medium with refractive index $n_{eff}$ = 1.518 matching the experiment. We include the isotropic, homogenous chirality by modifying the constitutive relations for the electric displacement field $\vec{D}$ and the magnetic induction $\vec{B}$ in the response of the chiral film and incorporate it into the COMSOL solver through the Pasteur parameter κ, as[33,50]:

$$\vec{D} = \varepsilon\vec{E} - \frac{i\kappa}{c}\vec{H} \qquad \text{Eq 2-1}$$
$$\vec{B} = \mu\vec{H} + \frac{i\kappa}{c}\vec{E} \qquad \text{Eq 2-2}$$

with $\varepsilon$ being the permittivity, $\mu$ the permeability, $\vec{H}$ the magnetic field, and $\vec{E}$ the electric field.[33] The complex chirality parameter $\kappa$ describes the molecular chiroptical response, modeled by a frequency-dependent Lorentzian function, associated with a molecular electronic transition (See Methods for details)[33].

**Figure 4**a shows the average absorption of LCP and RCP light for a bare chiral film (red line) and that of a bare array of AuNPs for a selected lattice spacing of $D_{squ}$ = 425 nm (black line). We assume a single *molecular* resonance at $\lambda_0 = 400$ nm. The absorption spectra of the bare AuNP array shows a LSPR peak around 570 nm, as well as Rayleigh Anomalies (RAs) and SLRs, marked with arrows. For this lattice parameter, we obtain the <1,0> RA to be at RA = $n_{eff}$ · $D_{squ}$ = 646 nm, and the <1,1> RA at 457 nm. The <1,0> RA couples with the LSPR, resulting in a pronounced SLR at around 660 nm. The chiroptical response of both reference systems is shown in Figure 4b. As expected, no chiral activity is observed for the bare particle array (black line), whereas the chiral film (red line) shows a bisignated (i.e. having a negative and positive peak around the resonance position[64]) Lorentzian-like peak centered at 400 nm.

We analyze the absorbance and chirality of the hybrid system in Figure 4c and d, respectively. Similar to the experimental case, their average absorption spectra (blue and yellow lines) appear as additions of the spectral features of the chiral film and the AuNP array. All resonances can be clearly identified, and their near-fields show the expected localized (LSPR) and delocalized nature (SLR) (Figure 4e). In contrast, the differential absorbance, CD, of the hybrid system in Figure 4d shows clear features associated with the all plasmonic resonances of the achiral AuNPs (Figure 4c), not found in the spectra of the individual components (Figure 4b). In particular, beside a Lorentzian-like feature at the spectral position of the LSPR centered around 570 nm, distinctive and pronounced resonances around the spectral positions of the RA and SLR are observed.

Moreover, the chiral fingerprint of the molecular film seems to be modified by the influence of the plasmonic modes of the AuNP array. As in the experimental case, we hypothesize that the original shape of the molecular signal is not preserved because plasmonic AuNPs are not helicity preserving.[48]

All spectral features of the experimental hybrid system (**Figure 3**) are thus reproduced by the simulations (**Figure 4**). Our simulations are also consistent with recent theoretical work[50,65–67] and delocalised chiral fields of *intrinsic* chiral lattice modes[62] recently used for chiral sensing.[68]

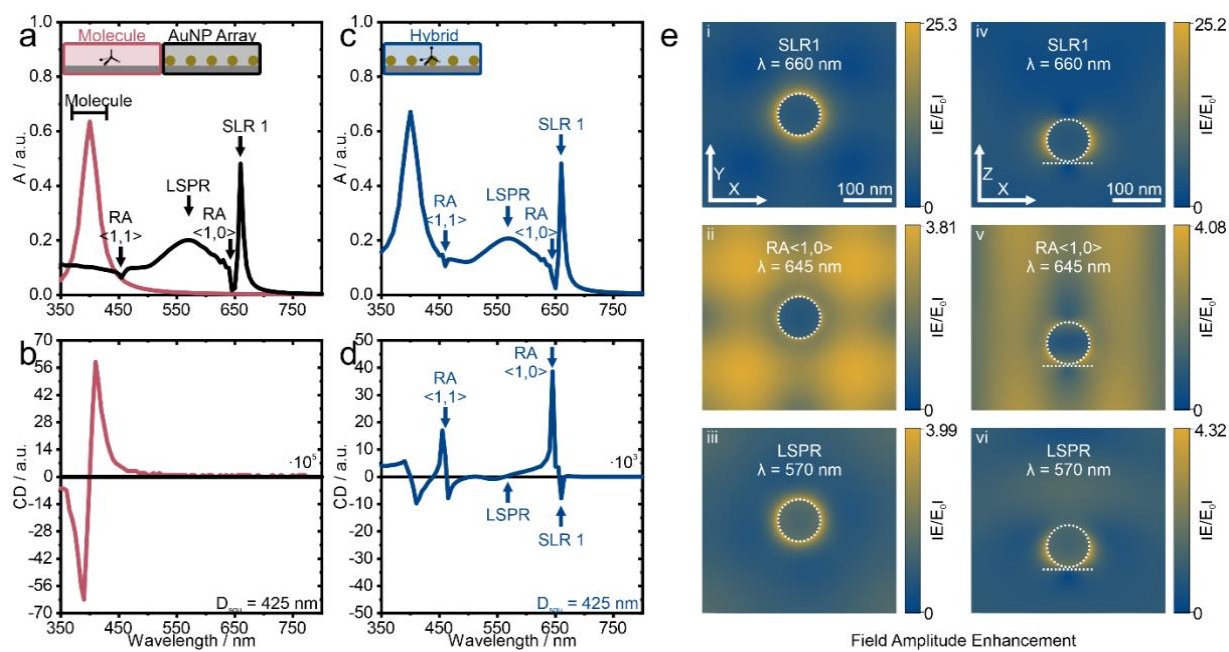

**Figure 4. Simulated chiroptical response of arrays of spherical AuNPs with and without the presence of chiral molecules.** a) Calculated mean absorbance of randomly oriented molecules in a chiral film for right and left circularly polarized (red line), and for the achiral square array of AuNP spheres (D = 90 nm) embedded in a homogeneous achiral medium with $n_{eff}$ = 1.518 (black line). The molecular response shows a pronounced resonance peak at 400 nm, and the particles array exhibits a surface lattice resonance (SLR) at 660 nm, localized surface plasmon resonance (LSPR) at 570 nm and Rayleigh anomalies (RA) at 426 nm and 645 nm. b) The differential absorption (CD) shows a chiral signature for the molecular resonance, while the plain plasmonic arrays remains achiral (flat line at CD = 0). c) The mean absorption for the hybrid system combining chiral film (blue line) exhibits all spectral features of the individual components. d) The circular dichroism (CD) shows strong absorption differences at the plasmonic resonances (LSPR, SLR1) and the Rayleigh anomalies. e) Amplitude of the electric field enhancement at the wavelengths of interest for the hybrid system from the top (i-iii) and side (iv-vi) view. Cuts are through the center of the sphere, dashed lines indicate the position of the AuNP and the substrate/chiral film.

We now systematically vary the lattice parameters and investigate the occurrence and efficiency of the chirality transfer to the SLR modes. In **Figure 5**a, we show the simulated absorbance for the different lattices in the presence of the molecular chiral film (hybrid system). As expected, the system shows multiple higher-order RAs for large enough lattice parameters ($D_{squ}$) appearing as dips in the absorption spectra. To quantify the chiroptical effect on these resonances, we show the differential absorbance (CD) for selected values of $D_{squ}$ in Figure 5b (full set in Supporting Figure 6). Similar to the case of $D_{squ}$ = 425 nm (**Figure 4**c,d) we observe strong induced CD signatures around the <1,0> RA. This spectral shift demonstrates that the induced chirality mechanism results from the lattice modes interacting with the molecules. This chirality transfer can even be extended to higher order modes (<1,1>, <0,2>, <2,1>, <2,2>) for very large spacing $D_{squ}$. In Figure 5c, we plot the expected positions of the RAs (solid lines) together with the points of maximum chiral signal ($CD_{min/max}$). A closer look reveals the induced CD appears with a typical bisignated shape around the RA, i.e. resulting in a minima and maxima around the resonance position. To extract the maximum, we used the strongest of both peaks. As a result, we see that the optical activity coincides with the RAs for all lattice parameters. As expected for such large particle-separation, the chiral signal at the position of the LSPRs is fairly weak compared to the values around the RAs (**Figure 5**b). To emphasize this effect more clearly, in Figure 5d we show the CD *contrast,* defined as the difference between the CD maxima and minima at the position of a RA ($|CD_{max} - CD_{min}|$). This contrast exposes an effective and relatively large chirality transfer to the RAs through their coupling with the molecular CD. In spite of a smaller detuning to the molecular response, the LSPRs do not show such an effective chirality transfer for large AuNP separations (green squares in **Figure 5**,d), therefore, one can conclude that RAs are more effective in the process of induced chirality transfer compared to the LSPRs appearing.

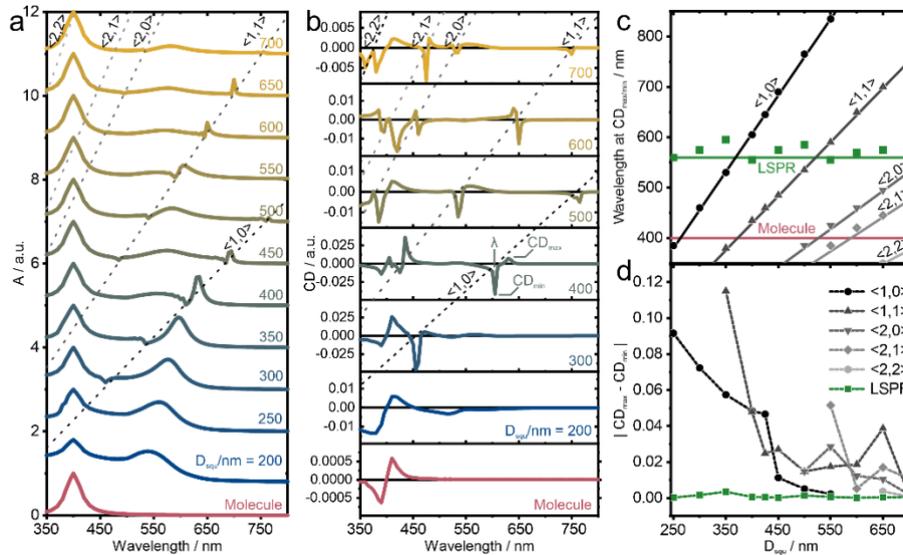

**Figure 5. Simulated absorbance and differential absorbance for different lattice spacings of the hybrid system.** a) Absorbance of randomly oriented molecules in a 200 nm thickness chiral film (bottom) for hybrid systems with different lattice parameter $D_{squ}$, embedded in a matrix with $n_{eff}$ = 1.518 illuminated by circularly polarized light. Dashed lines crossing diagonally are guides to the eye to indicate in-plane Rayleigh anomalies (RAs) of different order. b) Differential absorbance (CD) for selected hybrid systems. c) Rayleigh anomalies (RAs) as function of the lattice parameter $D_{squ}$. Solid diagonal lines represent the different in-plane diffractive orders, as labeled. The data points are the peak positions extracted from the individual CD spectra. Horizontal lines mark the spectral position of the LSPR and the molecular resonance. d) CD contrast, defined as the difference between the maxima and the minima of the calculated differential absorbance at the different RAs and at the LSPR.

## 2.4 Insights into the mechanism of chirality transfer

Electromagnetic interactions are known to be relevant in chirality transfer in hybrid systems composed of chiral molecules and plasmonic nanoparticles.[66,69,70] In particular, the electromagnetic coupling between the nanoparticles and the chiral molecules can induce differential dissipative currents in the plasmonic particles.[71] These currents can induce a prominent CD response at the plasmonic resonance frequencies even if the plasmonic system is achiral, a phenomenon known as the plasmon-induced CD effect. The full wave simulations of the array of gold nanospheres surrounded by a chiral film shown in **Figure 4** naturally capture this effect. With the aim of qualitatively understanding the phenomenon of chirality transfer to the lattice modes, we developed a semi-analytical model based on the coupling of a dipolar chiral molecule to a generic optical resonator described through a dipolar polarizability which mimics the spectral properties of the plasmonic array under study (see Supporting Information). The interaction between the chiral molecule and the plasmonic resonator is described within the Dyadic Green's tensor formalism under the coupled dipole approximation.[69]

**Figure 6** summarizes the results of this phenomenological model. Figure 6a shows both the extinction cross section of a chiral molecule (red line), and the polarizability mimicking the optical response of the nanoparticle array (black line). The extinction cross section of the molecule presents a Lorentzian line shape resonance at $\lambda = 400$ nm similar to the simulated molecular resonance in Figure 4a (red line). The optical response of the plasmonic array is characterized by three Lorentzian resonances, one for each lattice resonance and another one for the LSPR. This reproduces the LSPR around 550 nm, the SLR at 650 nm and RAs at 460 nm and 645 nm. Figure 6c presents the extinction cross section of the hybrid system. Since we consider the resonator strength of a single molecule, the molecular polarizability is substantially weaker than the plasmonic resonances and is thus not resolved in the spectra. Also note that different positions and orientations of the molecular dipole were averaged to capture the effect of the random distribution of molecules in the experimental system. The CD spectra for both the bare molecule and the plasmonic array are shown in Figure 6b. As expected, only a feature corresponding to the chiral molecular resonance is observed. The CD spectrum of the hybrid system (Figure 6d) shows clear chiral signals at the spectral positions of all plasmonic modes, in agreement with experiment (Figure 3) and full numerical simulations (Figure 4). It is remarkable that the polarizability of a single molecule, hardly detectable in direct extinction spectra, is able to produce such a large chiral signature at other, tunable spectral regions. The qualitative agreement provided by the coupled-dipole model points towards the basic mechanism of chirality transfer. A molecular chiral signal, even if largely detuned from the plasmonic resonance, is able to transfer its chirality via the overlapping of the spectral tail of the molecular signal with the plasmonic resonance. This overlap, even if small, is actually the only form of coupling considered in our model. The coupling is particularly pronounced if the plasmonic resonance exhibits spectrally narrow resonances, as it occurs in lattice modes. This is also the reason why the chirality transfer to the LSPR, even though spectrally less detuned, is less efficient. Note that considering the chiral molecule under the influence of the near-field created by the plasmonic array, without considering the coupling, does not reproduce the observed chirality transfer (details in Supporting Information).

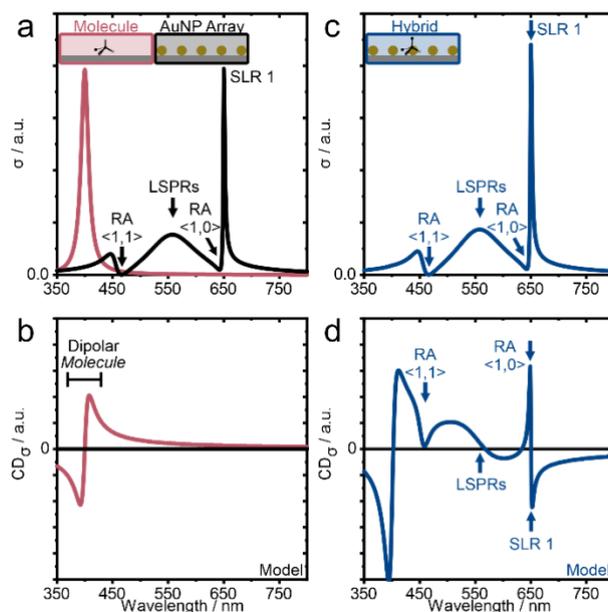

**Figure 6. Phenomenological model of a dipolar chiral molecule coupled to an optical resonator with spectral properties similar to the plasmonic array.** a) Calculated extinction cross section of the dipolar chiral molecule and the polarizability mimicking the optical response of the plasmonic nanoparticle array. b) Calculated extinction CD spectra of the dipolar chiral molecule and the plasmonic nanoparticle array. c) Calculated extinction cross section of the hybrid system consisting of a single molecular dipole and the plasmonic array. Note that the single molecular resonance is absent in the spectra due to the weak oscillator strength. d) Calculated extinction CD spectra of the hybrid system spatially averaged for all possible molecular locations.

# 3    Conclusion

In summary, we experimentally demonstrate a molecular chirality transfer to plasmonic lattice resonances. In the presence of a chiral molecular film, achiral AuNP arrays exhibit clear CD peaks at the position of their SLRs. Numerical simulations using AuNPs with systematically varying lattice constants embedded in a chiral film corroborate the experimental results and show chiral signals that spectrally coincide with the RAs and SLRs. A phenomenological model based on coupled dipoles including full electromagnetic interactions identifies the electromagnetic coupling, through the spectral overlap, as responsible for the chirality transfer.

Chirality transfer to SLRs offers an exciting possibility to tailor the spectral position of a detected chiral signal simply by changing the interparticle distance. Hence, such systems may be tuned to specific wavelengths to match desired light sources and detectors with optimised single-wavelength polarizing components. This makes such a platform suitable for in-line quality control measurements in (pharmaceutic) production sites, chromatographic sensors, or in highly sensitive microfluidic devices. More sophisticated fabrication processes, providing SLRs with very high Q-factors[56] may significantly enhance the induced CD signal. Local functionalization strategies[72] or specifically targeted molecular motives[73,74] may improve sensitivity and selectivity of the chiral response, and possibly provide control of the orientation of the chiral molecules with respect to the particle array to further enhance the chirality transfer.

# 4 Experimental Methods

## 4.1 Materials

All chemicals were used as received without further purification unless specified otherwise. Chemicals for synthesis, assembly, and sensing were all purchased by Sigma-Aldrich or Carl Roth in synthesis or chromatography grade unless specified other. Ultrapure water was from a Milli-Q system (18.2 MΩcm, Elga PURELAB Flex). All substrates were glass substrates (microscopy slides, Objektträger) from Menzel/Carl Roth. Immersion Oil (n = 1.518) was purchased from Cargille (Type LDF, very low autofluorescence). Plasmonic Au-PNIPAM core−shell microgels were synthesised according to the protocol published elsewhere.[58,59]

## 4.2 Induced Chirality Samples

The assembly at the air-water interface and synthesis of the Au-PNIPAM core−shell microgels ($d_{Core}$ ~ 99 nm) is based on established protocols from our groups and is described in more detail in Ref.[58,59] The polymeric shells were removed by oxygen plasma treatment (4 sccm $O_2$, 100 W, 5-10 min.) without affecting the gold nanoparticle assembly. To ensure comparability with literature, the functionalization with a chiral test molecule and a homogenous refractive index environment, samples were prepared following a modified literature protocol.[35] Poly(methyl methacrylate) (PMMA, 5 wt.%) was dissolved in dimethyl sulfoxide at 60° C under constant stirring (350 rpm) overnight. No filtering or purification was applied at any point. After cooling to ambient, riboflavin (40 mM) was added and stirred in a dark place for 4 hours. The slim substrates containing AuNP arrays were attached to a sacrificial cover slip (26x26 mm²) with double-sided carbon tape for spin-coating. The polymeric solutions (400 µl) were spin-coated (30s, 3400 rpm, 400 µl) on the mounted samples within reduced ambient light (dark room). Samples were placed in a desiccator to remove the solvent by vacuum (~ 60 min). The backsides of the samples were cleaned with EtOH and Kimwipes prior to measuring CD spectra. A small drop of immersion oil was sandwiched between the substrate and a cover slip (same as substrate here, see **Supporting Figure 2** and **Supporting Figure 3**).

## 4.3 Characterisation

### 4.3.1 Microscopy

SEM images were taken using a GeminiSEM 500 (Zeiss, Germany) at 1 kV using the in-lens detector. AFM images (original size 10x10 µm$^2$) were measured with a Nanowizard 4 (JPK Instruments) in intermittent contact mode using OTESPA-R3 tips (Bruker) and analysed with ImageJ software.

### 4.3.2 Spectroscopy

UV-Vis-NIR spectra (only in SI) were measured using a conventional spectrometer (Lambda 950, Perkin-Elmer) with a Glan Thompson polarizer drive (Perkin-Elmer, B050-5284), typically in the range between 400 and 2500 nm in 2 nm steps and 0.48 s integration time. Molecular CD and absorption spectra for circularly polarized light with high sensitivity were recorded using a commercial CD spectrometer (Jasco J815) with 1 s integration time at 50 nm/min. The substrates were attached in a custom build substrate holder (**Supporting Figure 3**), allowing normal illumination and fitting into the temperature-controlled ($N_2$ flow, 21 °C) cuvette compartment.

### 4.4 Electrodynamic simulations

Differential absorbance $\Delta A = A_{RCP} - A_{LCP}$ (RCP and LCP referred to right- and left- circularly polarized light) shown throughout the paper and representing the optical response of the chiral slab/gold NP system upon interaction with circularly polarized light at normal incidence were obtained by numerically solving the full set of Maxwell's equations for a bi-anisotropic media by performing the Finite Element Method (FEM) implemented in the commercial software COMSOL Multiphysics[63] using the radio frequency module (RF) in the frequency domain.

The complete system consisting of an array of gold NPs and a chiral film containing chiral molecules (riboflavin) was modelled using the lumped port formulation and imposing periodic (Floquet) boundary conditions at the in-plane directions of a squared unit cell (x- and y-axis) and scattering boundary conditions in the out-of-plane direction (z-axis). To represent the experimental conditions, a single gold spherical nanoparticle (R = 45 nm) was assumed to be deposited at the bottom part of a 200 nm thickness slab representing the chiral film. Both, gold NPs and the chiral slab were placed in the middle of a sufficiently long (~1-2 µm) z-directed environment, representing the simulation unit cell. Accordingly, the physical domains were placed in regular square array arrangements with corresponding lattice parameters. Finally, perfectly matched layers (PMLs) with a thicknesses $T_{PML}$=500 nm were placed at the bottom and top parts of the simulation unit in addition to the scattering boundary conditions in the out-of-plane axis in order to avoid spurious reflections coming from the interfaces along this direction.

All domains in the simulation box were meshed by using tetrahedral elements maintaining a maximum element size mesh below λ/10 where λ is the wavelength of the incident light. For the elements corresponding to both the sphere and the slab domains, the size was ten times finer than the largest element size until convergence. A homogeneous refractive index $n_{eff} = 1.518$ was used for all the simulation box, except for the gold sphere and for the chiral film. For the former, the optical functions were taken from experimental data available in Ref. [75] and for the latter, we used the theoretical scheme followed, among others, in Refs. [33,40,50]. The dispersive chirality parameter $\kappa(\omega)$ of the chiral film used in the electromagnetic calculations was modelled by a frequency-dependent Lorentzian function $\kappa(\omega) = A\big(f(\omega) + ig(\omega)\big)$ associated with molecular electronic transitions with $f(\omega)$ and $g(\omega)$ given, respectively, by $f(\omega) = \frac{\omega_0^2 - \omega^2}{(\omega_0^2 - \omega^2)^2 + \omega^2 \Gamma^2}$ and $g(\omega) = \frac{\omega \Gamma}{(\omega_0^2 - \omega^2)^2 + \omega^2 \Gamma^2}$. The parameter A relates to the density of chiral molecules, $\omega_0$ represents the angular frequency of resonance and $\Gamma$ is the broadening of this resonance. All numerical calculations used a single molecular resonance, numerically set at $\omega_0 = 3.1$ eV ($\lambda_0 = 400$ nm) with a damping parameter $\Gamma = 0.2$ eV.

## 5 Author contribution

ESAG developed the idea and experimental concept with the help of MK and NV; ESAG and NV conceived the study. EP self-assembled the AuNP arrays and measured AFM; ESAG fabricated the final samples, performed electron microscopy, photography, chiroptical investigations, performed renderings, compiled figures and developed a preliminary theory; MZH performed the numerical calculations and analyzed and discussed the results with J.A. and AGE; MZH, AGE and JA applied the chirality transfer theory; all authors discussed the results and contributed to the final version of the manuscript; ESAG, MZH and NV wrote the manuscript with corrections from all authors. NV, MK and JA supervised the project.

## 6 Acknowledgement

We thank Markus Russ for the help with the sample holder and its fabrication. We thank Florian Golombek, Kathrin Castiglione, Benedikt Schmid, and Yves Muller for providing access to their CD spectrometers and their very kind hospitality. We thank Lisa V. Poulikakos and Wim Noorduin for their stimulating discussions and encouragement.
This project received funding from the European Union's Horizon 2020 research and innovation programme under grant agreement No 861950, project POSEIDON. N.V. acknowledges support from the Interdisciplinary Center for Functional Particle System (FPS) at FAU Erlangen-Nürnberg. A.G.E. acknowledges support from the Spanish Ministerio de Ciencia e Innovación (PID2019-109905GA-C2), from programa Red Guipuzcoana de Ciencia, Tecnología e Innovación 2021 (Grant Nr. 2021-CIEN-000070-01) and from Eusko Jaurlaritza Elkartek Program KK-2021/00082 and from IKUR Strategy under the collaboration agreement between Ikerbasque Foundation and DIPC on behalf of the Department of Education of the Basque Government, Programa de ayudas de apoyo a los agentes de la Red Vasca de Ciencia, tecnología e Innovación acreditados en la categoría de Centros de Investigación Básica y de excelencia (Programa BERC) Departamento de Universidades e Investigación del Gobierno Vasco and the Centros Severo Ochoa AEI/CEX2018-000867-S from the Spanish Ministerio de Ciencia e Innovación. M.K. acknowledges the German Research Foundation (DFG) for funding under grant KA3880/6-1. D. F. acknowledges the Luxembourg National Research Fund (FNR), Project Reference 15688439.

Competing interests:
The authors declare that they have no competing interests.

## 7 Data availability

Raw data and images are accessible from the corresponding authors upon reasonable request.

# Supporting Information for: Molecular-Induced Chirality Transfer to Plasmonic Lattice Modes

E.S.A. Goerlitzer[1,§,*], M. Zapata-Herrera[2], E. Ponomareva[3], D. Feller[3], A. Garcia-Etxarri[4,5], M. Karg[3], J. Aizpurua[2,4,*], N. Vogel[1,*]

[1] Institute of Particle Technology, Friedrich-Alexander University Erlangen-Nürnberg, Cauerstraße 4, D-91058, Erlangen, Germany

[2] Materials Physics Center CSIC-UPV/EHU, Paseo Manuel de Lardizabal 5 20018, Donostia-San Sebastián, Spain

[3] Institut für Physikalische Chemie I: Kolloide und Nanooptik, Heinrich-Heine-Universität Düsseldorf, Universitätsstr. 1, Düsseldorf, D-40225 Germany

[4] Donostia International Physics Center (DIPC), Paseo Manuel de Lardizabal 4 20018, Donostia-San Sebastián, Spain

[5] IKERBASQUE, Basque Foundation for Science, Maria Diaz de Haro 3, 48013 Bilbao, Spain

[§] Current address: NanoPhotonics Centre, Cavendish Laboratory, University of Cambridge, Cambridge CB3 0HE, U.K.

* Corresponding authors: eric.goerlitzer@fau.de, aizpurua@ehu.eus, nicolas.vogel@fau.de

# S1. Schematic overview of molecular-induced chirality in plasmonic AuNPs and their arrays.

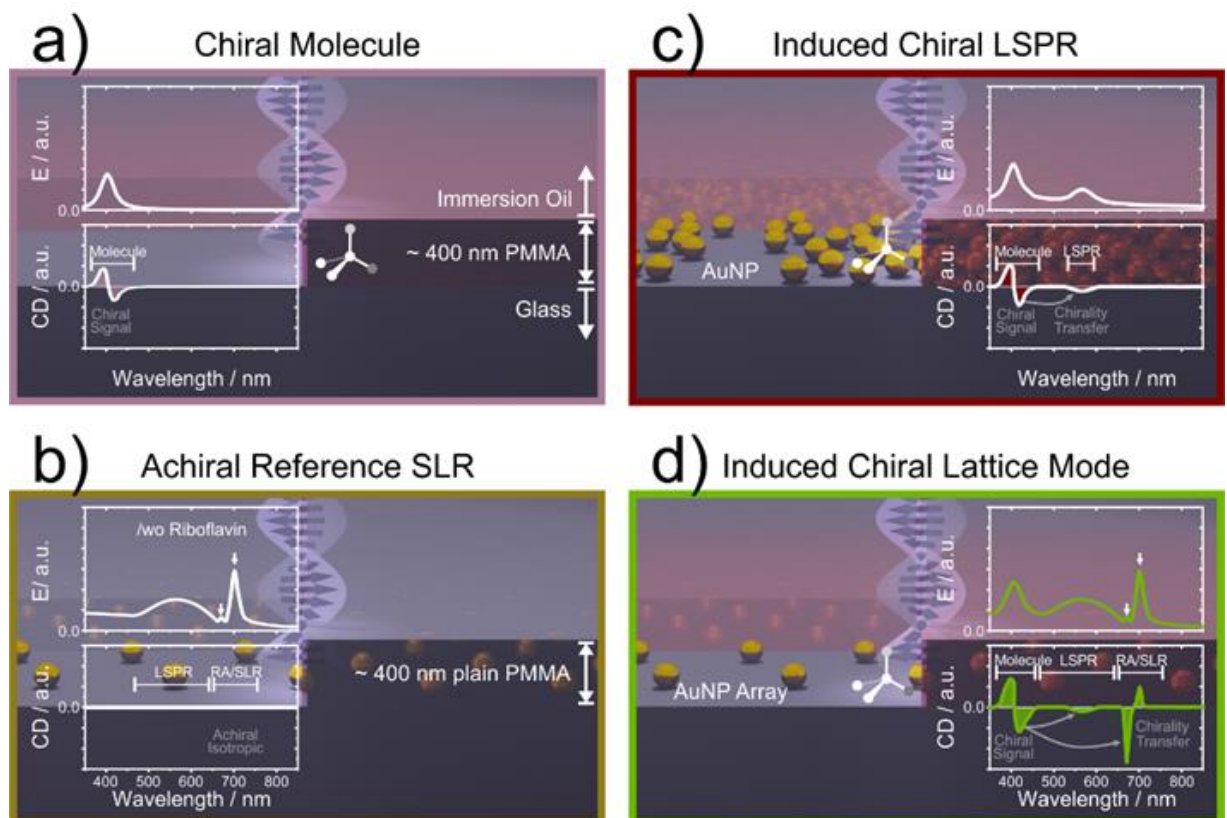

**Supporting Figure 1.** Schematic overview of molecular-induced chirality in plasmonic AuNPs and their arrays. a) The schematic chiroptical response of a slab of randomly oriented molecules. Molecules typically show an absorption in the UV region accompanied with a CD signal nearby b) Achiral AuNP embedded in an achiral homogenous RI environment show strong lattice modes (RA, SLR), but no signal in the circular dichroism. c) A hypothetic chiral film filled with randomly oriented molecules and AuNPs with no long-range order will show a chirality transfer from the molecular region to the LSPRs. d) If the AuNP have a long-range order, we suspect the chirality will be transferred in addition to LSPR also to the lattice modes, i.e. the RA and SLR of the system.

## S2. Impression of the sample fabrication and measurement.

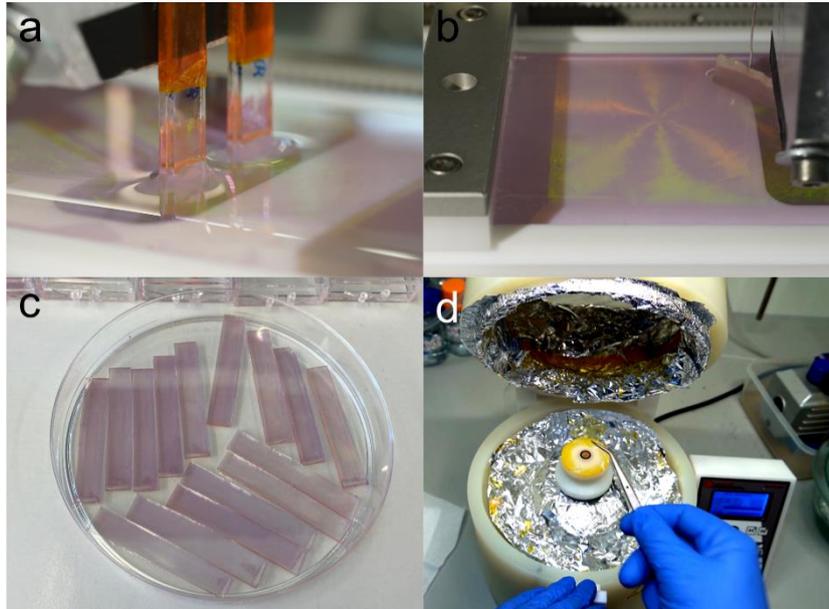

**Supporting Figure 2.** Impression of the sample fabrication and measurement. a) The 99 nm AuNP @ 367 nm PNiPAm form hexagonal arrays when assembled at the air-water interface, allowing the transfer to glass substrates. b) The ordering of the particles within the array is reflected by the appearance of Bragg modes, leading to the observable iridescent coloration. c) Photograph of typical samples after the self-assembly, with the polymeric shell (PNiPAm) still on top. d) After the shell is removed by oxygen plasma, we can use a spin-coater to add the (chiral) PMMA film on top, followed by vacuum drying (not shown here).

## S3. Custom build sample holder allowing to measure solid substrates in a commercial CD spectrometer.

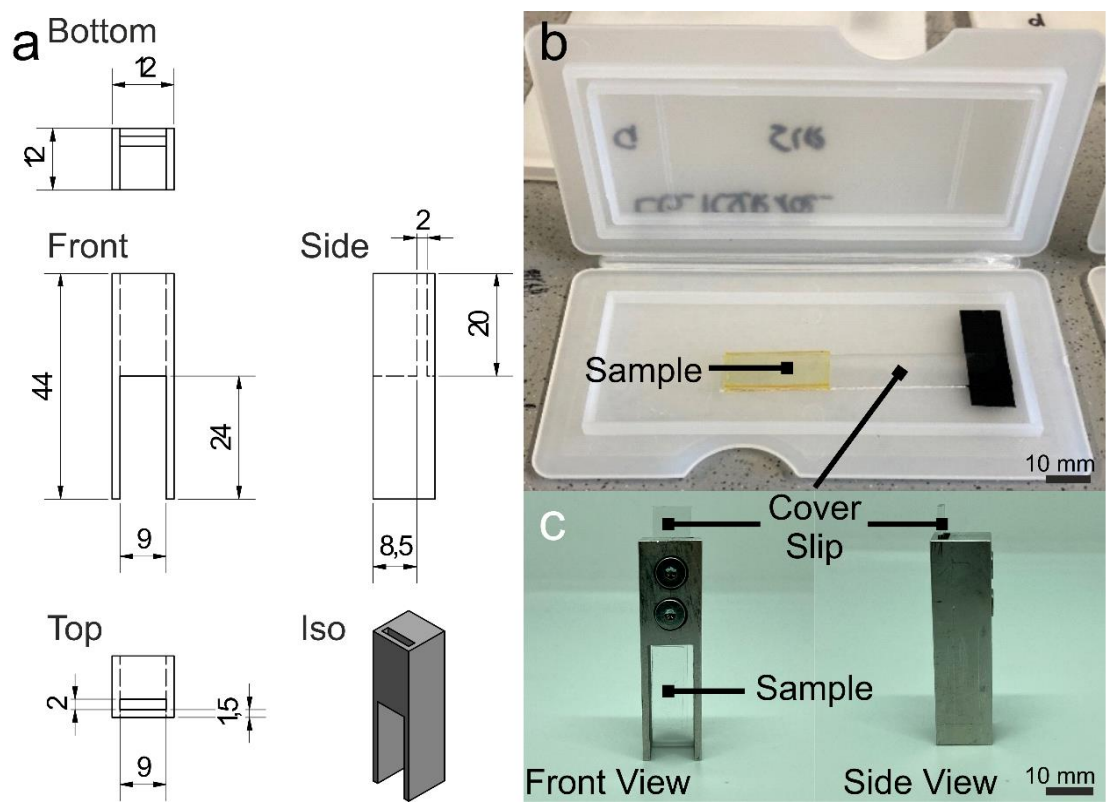

**Supporting Figure 3.** Custom build sample holder allowing measuring solid substrates in a commercial CD spectrometer (Jasco J815a), which is typically used to analyze liquid samples in a measuring cuvettes. a) Engineering drawing to build the sample holder, which was made form one solid aluminum block. Unit is mm. b) Final samples are sandwiched by immersion oil and a cover slip (i.e. the same glass as the substrate) to ensure a homogenous RI environment. c) The samples can be measured by putting the cover slip into the slit of the sample holder. The cover slit is fixed by screws, the sample is hold by capillary forces fixing the substrate in the illuminated area (bottom part). The entire sample holder fits into cuvette compartments.

## S4. Further supporting spectra of the investigated samples and additional samples, measured by the commercial CD spectrometer

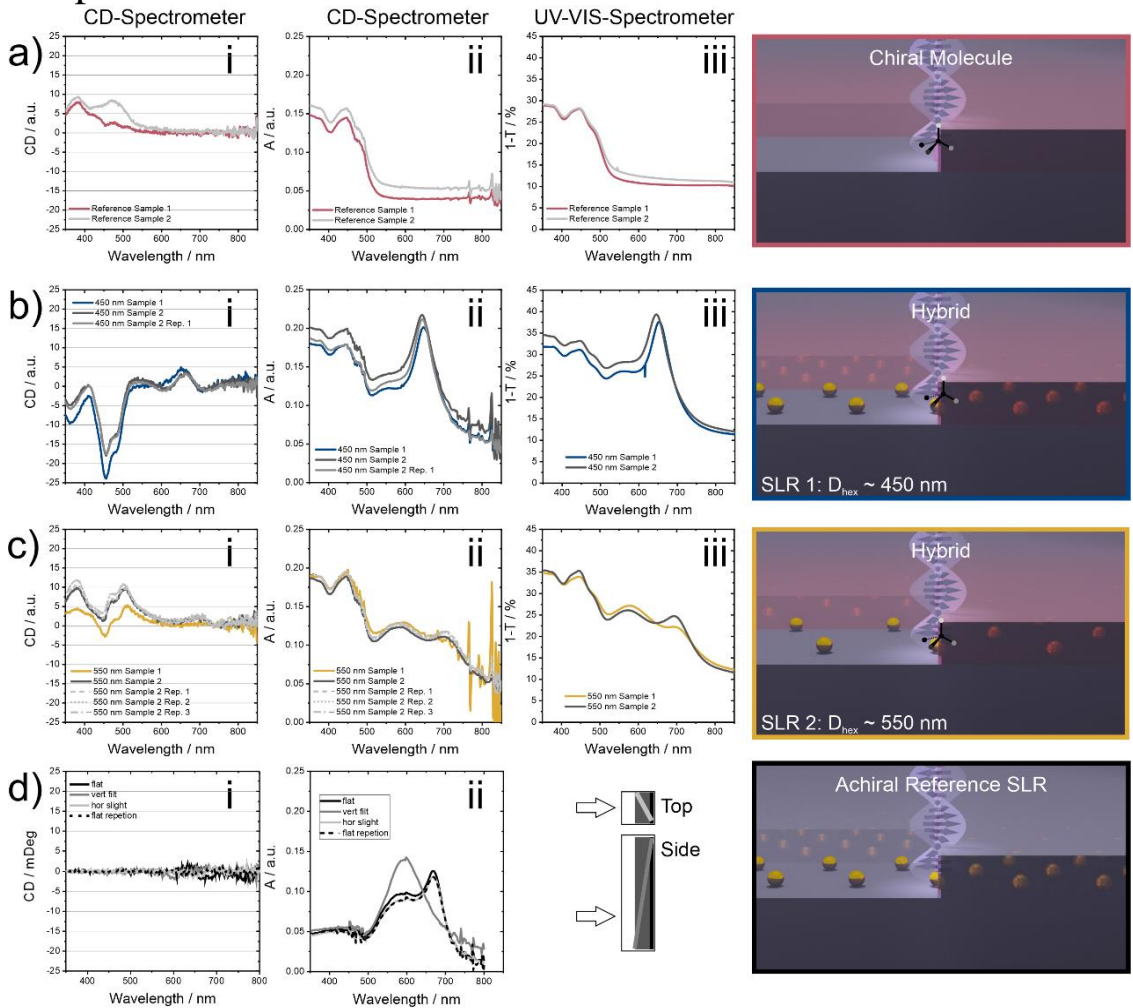

**Supporting Figure 4.** Further supporting spectra of the investigated samples and additional samples, measured by the commercial CD spectrometer (i: CD, ii: average A (LCP, RCP light) used in the main study, and with a commercial UV-VIS spectrometer (iii: 1-T, unpolarized light). a) Reference samples of only a film of PMMA with Riboflavin on top of a glass substrate (RI matched). b) Sample with AuNP arrays ($D_{hex}$ ~ 450 nm) and PMMA with Riboflavin. c) Sample with AuNP arrays ($D_{hex}$ ~ 450 nm) and PMMA with Riboflavin. Note, that a and b are different samples with similar geometry in b) and c) respectively. Samples used in the main text (b) were measured at least twice with referencing the device against air in-between. d) Sample with only AuNP arrays ($D_{hex}$ ~ 450 nm) and plain PMMA. Note, the tilted measurements (blue and pink line) were performed in a glass cuvette to allow vertical and horizontal tilting respectively. Tilting has no significant effect on the resulting CD spectrum as arrays of spherical AuNP particles are unsusceptible to extrinsic and intrinsic chirality.

## S5. Hexagonal and Square Lattices

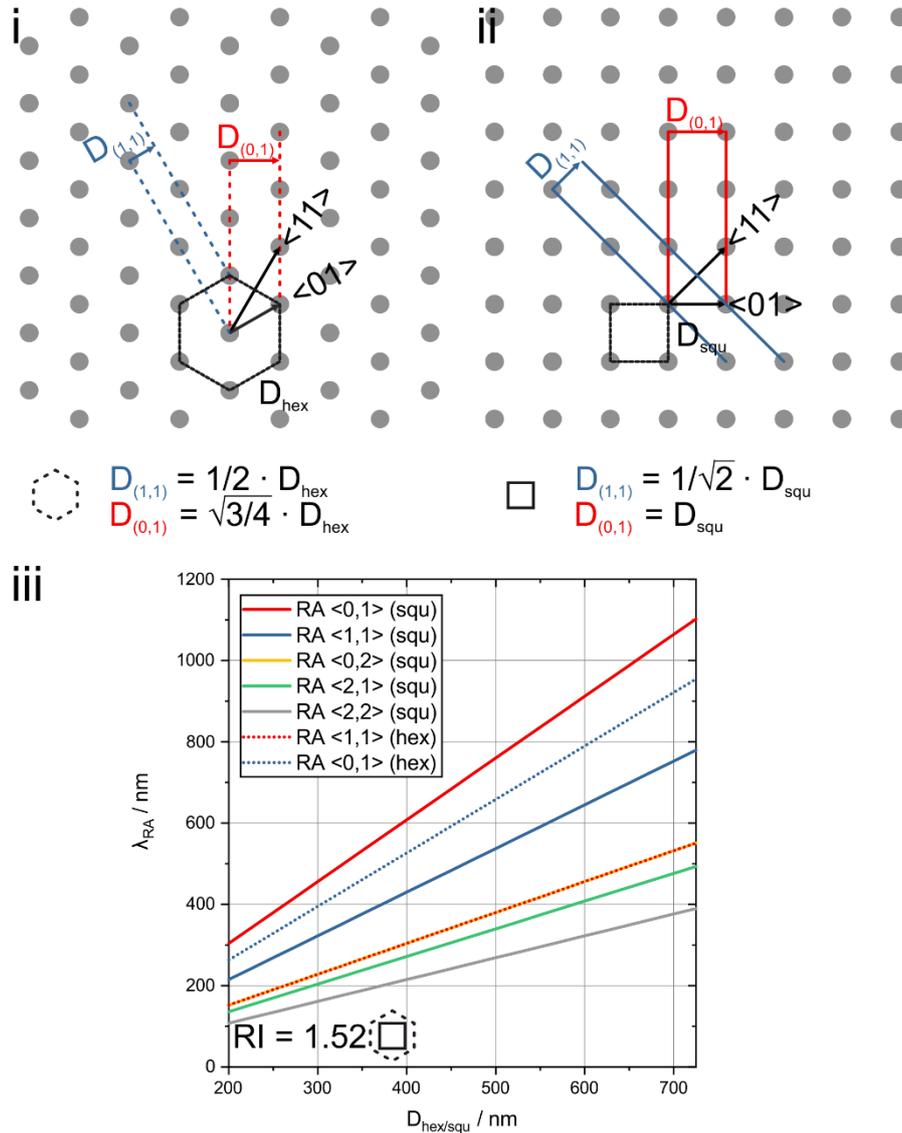

**Supporting Figure 5.** Crystal planes, directions and RAs in 2D hexagonal (i) and square (ii) arrays. The position of the RAs is determined by multiplying the refractive index (RI) with the interplanar distances (D(1,0) and D(1,1)). The latter, in turn, depend on the crystallographic type and their interparticle distances ($D_{hex}$ and $D_{squ}$). iii) The spectral position of RAs for different interparticle distances are calculated for square arrays for a RI of 1.518, as well as for a hexagonal array (dotted line).

## S6. Full set of chiroptical spectra as function of the lattice spacing

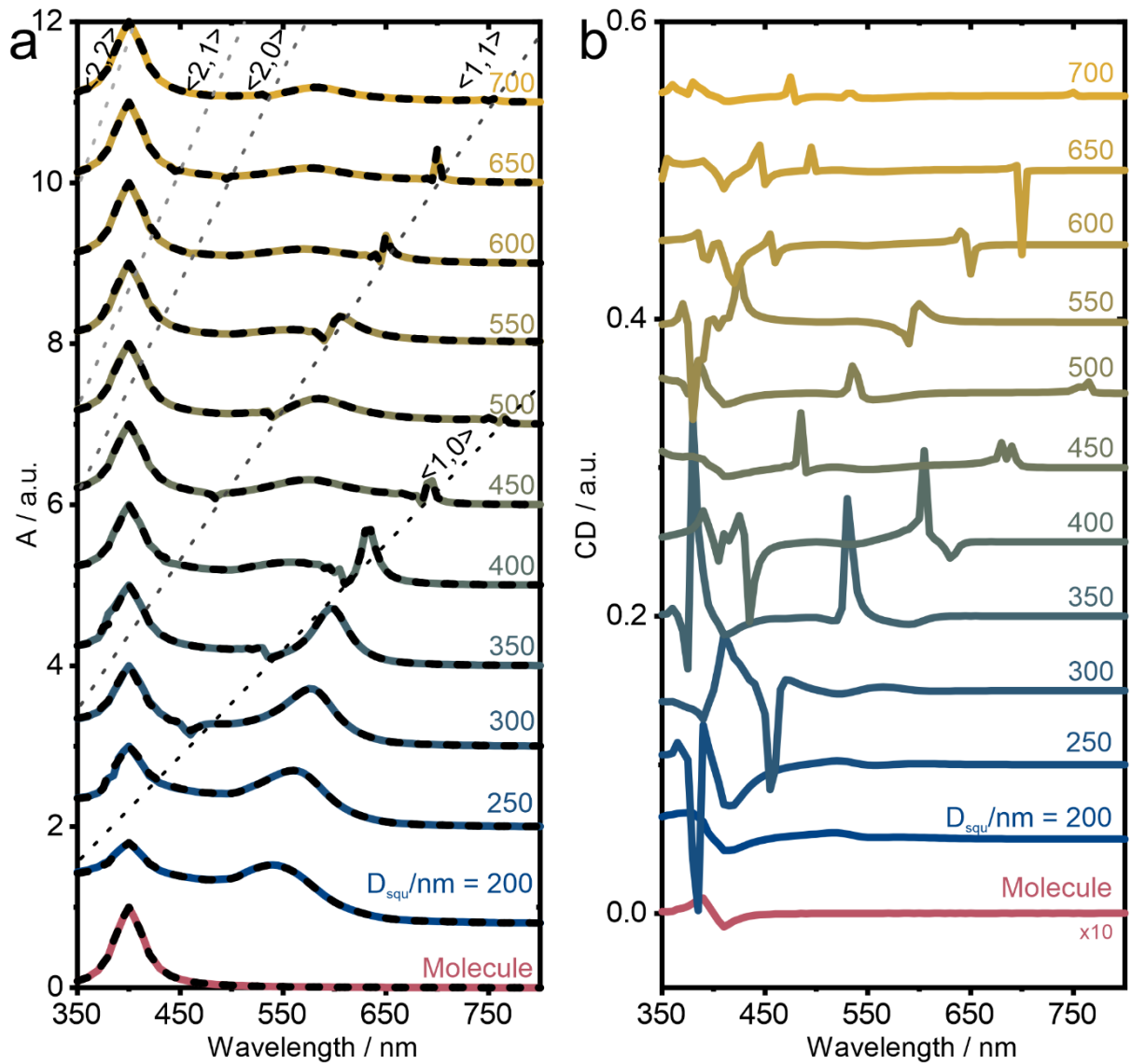

**Supporting Figure 6.** Chiroptical response for different lattice spacing. a) Calculated optical absorbance of randomly oriented molecules in a 200nm thickness chiral film (bottom) and hybrid system (AuNP D = 90 nm) with different lattice parameter $D_{squ}$, embedded in a matrix with a RI = 1.518 illuminated with right and left circularly polarized light (RCP, LCP; continuous and black dashed lines, respectively). Crossing lines are guides to the eye to indicate the corresponding different in-plane diffractive orders, i. e. the Rayleigh anomalies (RAs) accessible in the spectral range. b) Differential absorbance (CD) for the corresponding lattices.

## S7. Theory

One could tentatively think that the observed chirality transfer phenomenon can be related to an enhancement of the circular dichroism of the chiral molecules due to the intense near field distributions induced on the plasmonic lattice at the different resonances (**Figure 4**e-f). On a first approximation, assuming that the molecules cannot alter the optical response of the nanoparticles, the enhancement of molecular CD in complex non-chiral environments can be quantified through the electromagnetic density of chirality, normalized to that of a plain circular wave in free space ($C/C_{CLP}$).[1,2] It is well known that field-enhanced CD spectroscopy necessitates optical antennas supporting both electric and magnetic dipolar resonators. In the past, it has been shown that high index nanoparticles[1,3] and nanoparticle lattices[4,5] can sustain CD enhancement factors which preserve their sign all around the nanoparticles. In our system the local electromagnetic density of chirality can be substantially enhanced in certain regions of space (See **Supporting Figure 7**). Nevertheless, as predicted in Ref [1], due to the electric-dipolar nature of the plasmonic resonances excited here and the absence of magnetic-dipolar excitations, the lattice sustains positive and negative enhancement regions, and the overall CD enhancement cancels out when integrated along a volume surrounding the lattice.

The local density of optical chirality, $C(\vec{r})$, defined by [6,7]

$$C(\vec{r}) \equiv -\frac{\omega}{2c^2} \text{Im}[\vec{E}^*(\vec{r}) \cdot \vec{B}(\vec{r})], \quad (S7.1)$$

establishes the conditions for enhancing the CD signal when chiral molecules in the vicinity of a nonchiral optical resonator (a plasmonic nanoantenna array, in our case) interacts with the local field of the system. In Eq. S7.1, $\vec{E}$ and $\vec{B}$ represent the complex electric and magnetic fields, and $\omega$ and $c$ are the angular frequency and velocity of light in vacuum, respectively. Generally, a favourable contribution to the CD signal is because $C(\vec{r})$ preserves its sign in space when the nanostructure is illuminated by circularly polarized light (CPL) of a given handedness. Thus, the molecular CD signal can be, in principle, enhanced in all regions of space in the presence of optical antennas. In particular, in the presence of nonchiral antennas, we can normalize the expression given by eq. S7.1 to that of a plain circular wave in free space, $C_{cpl} = \pm \frac{\epsilon_0 \omega}{2c} E_0^2$ (where $E_0$ is the magnitude of the incident electric field and $\epsilon_0$ is the permittivity of free space). Thus, in these situations, a local CD enhancement factor $f_{CD}(\vec{r})$ can be defined as [1]

$$f_{CD}(\vec{r}) = \frac{C(\vec{r})}{|C_{cpl}|} = -\frac{Z_0}{E_0^2} \text{Im}[\vec{E}^*(\vec{r}) \cdot \vec{B}(\vec{r})] \quad (S7.2)$$

with $Z_0$ being the impedance of vacuum.

By integrating over the volume containing the chiral film without the non-chiral nanoparticles (normalized to the same volume $V_{film}$), we can obtain the total enhancement factor of the CD $f_{CD}^{tot}$, defined as

$$f_{CD}^{tot} = \frac{1}{V_{film}} \iiint f_{CD}(\vec{r}) dV \quad (S7.3)$$

Under the assumption that the electromagnetic coupling between the molecules and the plasmonic system is very weak, one can predict the CD signal of the hybrid chiral film-gold NP (CD$_{hyb}$) by multiplying $f_{CD}^{tot}$ (calculated for the gold NP array without the chiral film) and the CD signal (or differential absorbance ΔA) of the chiral film without NPs

$$CD_{hyb} = f_{CD}^{tot} * CD_{film}.$$

In the case of a gold sphere array (D = 90 nm, lattice parameter D$_{squ}$ = 425 nm) embedded in an effective media with n$_{eff}$=1.518, illuminated with CPL and neglecting the influence of the chiral film through electromagnetic coupling, we calculate the $f_{CD}(\vec{r})$ maps at the wavelengths of the SLR at 660 nm, the RA at 645 nm and the LSPR at 570 nm as they appear in Supporting Figure 7a.

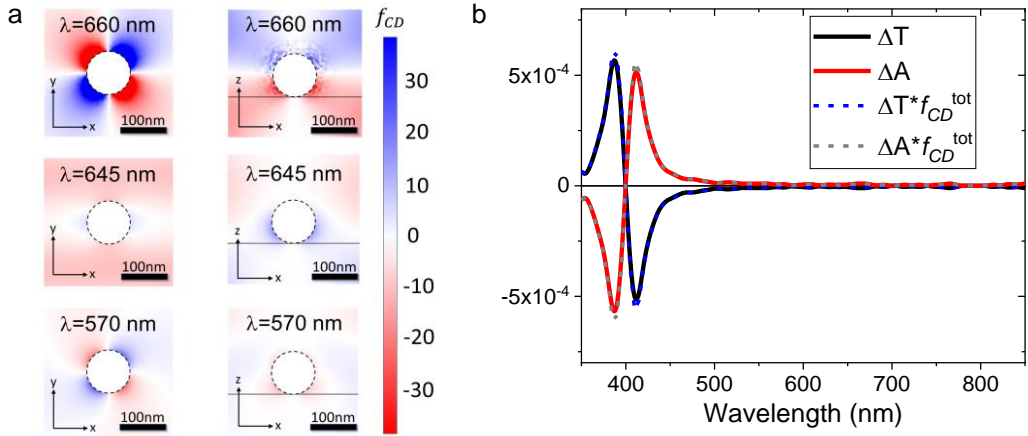

**Supporting Figure 7** a) Electromagnetic local density of chirality $f_{CD}(\vec{r})$ at interest wavelengths for the gold NP array embedded in a matrix with $n_{eff} = 1.518$ illuminated with right circularly polarized light. The AuNP diameter is $D = 90nm$ and the lattice spacing is $D_{squ} = 425nm$. b) Differential absorbance (Δ$A$, red continuous line) and transmittance (Δ$T$, black continuous line) calculated using COMSOL Multiphysics and via the total density of chirality $f_{CD}^{tot}$ multiplied by Δ$A$ (gray dashed lines) and Δ$T$ (blue dashed lines), respectively.

Because of the high values of $f_{CD}(\vec{r})$ around the gold nanoparticle, particularly for those at SLR, it is tempting to think that they may be responsible of the enhancement of the CD signal when the NP array is in interaction with the chiral film. Nevertheless, the differential absorbance (transmittance) calculated numerically for the chiral film and $f_{CD}^{tot} * CD_{film}$ (Supporting Figure 7b) turn out to be practically identical, thus indicating that the intense near field distributions around the plasmonic lattice are not inducing any enhancement of the molecular CD signal. In other words, despite of the high values of $f_{CD}(\vec{r})$, $f_{CD}^{tot} \approx 1$. The lattice sustains positive and negative enhancement values of the $f_{CD}(\vec{r})$ in symmetric regions, cancelling out any enhancement when integrated along a volume surrounding the lattice. Thus, the local density of chirality does not amplify the CD signal of the chiral film. This suggests that electromagnetic interactions between the chiral molecules and the plasmonic array play a crucial role in the experimentally observed chirality transfer phenomenon.

In the following section, we briefly sketch the field-mediated chirality information transfer theory in molecule-nanoparticle hybrids used to obtain the results presented in Figure 5 of the main text. This formalism, originally proposed in Ref.8 (see also[9,10]), unlike the surface-enhanced CD effect described in the previous paragraphs, accounts for the electromagnetic interactions between chiral molecules and optical resonators.

Our hybrid system, depicted in Supporting Figure 8, consists of an achiral array of plasmonic gold nanoparticles and a chiral film. In the following, we will try to capture the optical response of this system following a simple coupled dipole approach.

We will embed the optical response of the plasmonic lattice in a 6x6 dipolar polarizability.

$$\alpha_A = \varepsilon_0 \varepsilon_h \begin{pmatrix} \alpha_A^e & \alpha_A^{em} \\ \alpha_A^{me} & \alpha_A^m \end{pmatrix}$$

Where $\varepsilon_0$ is the vacuum permittivity and $\varepsilon_h$ is the permittivity of the embedding medium (in this case $\varepsilon_h = 1$). The isotropic electric polarizability $\alpha_A^e$, capturing the optical response of the plasmonic array, can be described by 3 Lorentzian resonances, one for each lattice resonance and another for the LSPR.

$$\alpha_A^e = V_A \left( A_1 e^{i\theta_1} \frac{\Gamma_1}{(\omega_1 - \omega) + i\Gamma_1} + A_2 e^{i\theta_2} \frac{\Gamma_2}{(\omega_2 - \omega) + i\Gamma_2} + A_3 e^{i\theta_3} \frac{\Gamma_3}{(\omega_3 - \omega) + i\Gamma_3} \right) \mathbb{I}_3$$

In our calculations we used we used the following values for the variables in order to describe the optical properties of the plasmonic lattice:

$V_A = 4/3 \pi a_A^3$
$a_A = 500 nm$
$nm = 10^{-9}$

$A_1 = 1.8$
$A_2 = 10$
$A_3 = 0.8$

$\theta_1 = \pi$
$\theta_2 = 0{,}875\pi$
$\theta_3 = 1{,}45\pi$

$\omega_1 = 2\pi \frac{c}{\lambda_1}$
$\omega_2 = 2\pi \frac{c}{\lambda_2}$
$\omega_3 = 2\pi \frac{c}{\lambda_3}$

$c = 3 * 10^8$

$\lambda_1 = 560 nm$
$\lambda_2 = 650 nm$
$\lambda_3 = 455 nm$

$\Gamma_1 = \omega_1/10$
$\Gamma_2 = \omega_2/400$
$\Gamma_3 = \omega_3/40$

$\mathbb{I}_3 = 3x3 \; identiy \; matrix$

Since the nanoparticle array is plasmonic and achiral, one can safely assume that $\alpha_A^m = 0 * \mathbb{I}_3$ and $\alpha_A^{em} = \alpha_A^{me} = 0 * \mathbb{I}_3$.

On the other hand, to analytically capture the optical response of the chiral film, we simplify the problem and consider the response of a single chiral molecule. To do so, we describe it as a chiral, magneto-electric dipole of polarizability:

with:
$$\alpha_B = \varepsilon_0 \varepsilon_h \begin{pmatrix} \alpha_B^e & \alpha_B^{em} \\ \alpha_B^{me} & \alpha_B^m \end{pmatrix}$$

$$\alpha_B^e = V_B \left( A_4 e^{i\theta_4} \frac{\Gamma_4}{(\omega_4 - \omega) + i\Gamma_4} \right) \mathbb{I}_3$$

$$\alpha_B^m = 0 * \mathbb{I}_3$$

And,
$$\alpha_B^{em} = -\alpha_B^{me} = V_B \left( A_5 e^{i\theta_5} \frac{\Gamma_5}{(\omega_5 - \omega) + i\Gamma_5} \right) \mathbb{I}_3$$

With the following values for the variables taken in order to describe the optical properties of the chiral molecule:

$V_A = 4/3 \pi a_B^3$
$a_B = 5 nm$

$A_4 = 1$
$A_5 = 10^{-3}$

$\theta_4 = \theta_5 = \pi$

$\omega_4 = 2\pi \frac{c}{\lambda_4}$
$\omega_5 = 2\pi \frac{c}{\lambda_5}$

$\lambda_4 = \lambda_5 = 400 nm$

$\Gamma_4 = \Gamma_5 = \omega_4/50$

The interaction between the plasmonic array (A-labeled in Supp. Fig.8) and the chiral molecule (B-labeled in Supp. Fig. 8), following reference[8], can be captured by the full description of the induced dipoles

$$\mathcal{P}_A^{L/R} = \alpha_A \begin{bmatrix} E_A^{tot} \\ Z_h H_A^{tot} \end{bmatrix} \quad \text{and} \quad \mathcal{P}_B^{L/R} = \alpha_B \begin{bmatrix} E_B^{tot} \\ Z_h H_B^{tot} \end{bmatrix}$$

Where $Z_h$ is the impedance of the medium. $E_A^{tot}$ and $H_A^{tot}$ are the total fields (scattered plus incident) at the position of the optical resonator, while $E_B^{tot}$ and $H_B^{tot}$ are the total fields at the position of the chiral molecule. L/R indicate left- and right- circularly polarized incident light respectively.

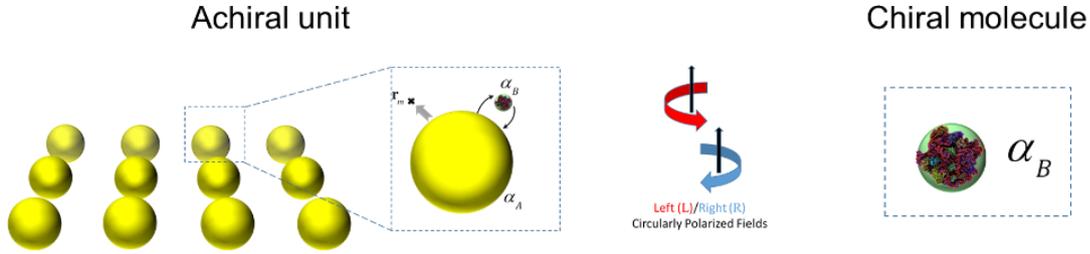

**Supporting Figure 8.** Schematic of the model describing the chiral molecule-nanoparticle hybrid system. $\alpha_A$ and $\alpha_B$ are the polarizability tensors describing the plasmonic nanoparticle array and the chiral molecule, respectively.

$\mathcal{P}_A^{L/R}$ and $\mathcal{P}_B^{L/R}$ need to be calculated self-consistently. Since the fields scattered by the chiral molecule will contribute to the total fields at the position of the optical resonator and vice-versa.

These coupled dipole equations are given by

$$\mathcal{P}_A^{L/R} = \varepsilon_0 \varepsilon_m \boldsymbol{\alpha}_A \left( \boldsymbol{\Psi}_0^{L/R}(\mathbf{r}_A) + \frac{k^2}{\varepsilon_0 \varepsilon_m} \mathcal{G}_0(\mathbf{r}_A, \mathbf{r}_B) \mathcal{P}_B^{L/R} \right) \quad (S8.1)$$

$$\mathcal{P}_B^{L/R} = \varepsilon_0 \varepsilon_m \boldsymbol{\alpha}_B \left( \boldsymbol{\Psi}_0^{L/R}(\mathbf{r}_B) + \frac{k^2}{\varepsilon_0 \varepsilon_m} \mathcal{G}_0(\mathbf{r}_B, \mathbf{r}_A) \mathcal{P}_A^{L/R} \right). \quad (S8.2)$$

In Eqs. S8.1-2, $\boldsymbol{\Psi}_0^{L/R}(\mathbf{r}_A) = \begin{bmatrix} E_A^{inc} \\ Z_h H_A^{inc} \end{bmatrix}$ corresponds to the incident electromagnetic field at the position of the optical resonator and $\boldsymbol{\Psi}_0^{L/R}(\mathbf{r}_B) = \begin{bmatrix} E_B^{inc} \\ Z_h H_B^{inc} \end{bmatrix}$ to the incident electromagnetic field at the position of the chiral molecule. $\mathcal{G}_0(\mathbf{r}_i, \mathbf{r}_j)$ is the free space Green's dyadic, [8] $k$ is the wave number of light in the medium.

Once the coupled set of equations (S8.1-2) are solved self-consistently, the extinction cross section of the coupled hybrid system can be obtained through the expression:

$$\sigma_{ext}^{L/R} = \frac{k}{\varepsilon_0 \varepsilon_m \left| E_0^{L/R} \right|^2} \mathrm{Im} \{ \boldsymbol{\Psi}_0^{*L/R}(\mathbf{r}_A) \mathcal{P}_A^{L/R} + \boldsymbol{\Psi}_0^{*L/R}(\mathbf{r}_B) \mathcal{P}_B^{L/R} \} \quad (S8.3)$$

In Figure 6 of the main text, the differential extinction cross section (Fig. 6d) is obtained from the contributions of Left- and Right- Circularly polarized incident light by

$$\Delta \sigma_{ext, \ A,B} = \sigma_{ext}^L - \sigma_{ext}^R \quad (S8.4).$$

The results in Figure 6 prove that chiral molecular resonance, even if largely detuned from the plasmonic resonance, is able transfer its chirality via the overlapping of the spectral tail of the molecular signal with the plasmonic resonance. Note, that to capture this chirality transfer mechanism it is essential to consider the full electromagnetic coupling between the chiral molecule and the optical resonator. Formalisms which exclude this coupling (see Supporting Figure 7) are unable this chirality transfer effect.

## S8. SI References